\documentclass[twocolumn,aps,prl,superscriptaddress,nofootinbib]{revtex4}

\usepackage{amssymb,xcolor}
\usepackage{amsmath}
\usepackage{amsbsy}
\usepackage{amsthm}
\usepackage{bbm}
\usepackage[dvips]{graphicx}

\newcommand{\beq}{\begin{equation}}
\newcommand{\eeq}{\end{equation}}

\newcommand{\bra}[1]{\langle#1|}
\newcommand{\proj}[1]{|#1\rangle\langle#1|}
\newcommand{\ket}[1]{|#1\rangle}

\newcommand{\id}{\mathbbm{1}}
\newcommand{\R}{\mathbbm{R}}

\newcommand{\tr}{\textnormal{Tr}}

\newcommand{\dicke}{\ket{D^{(ph)}_{4}}}
\newcommand{\bradicke}{\bra{D^{(ph)}_{4}}}

\begin{document}
	
	\title{Hyperentangled mixed phased Dicke states: optical design and 
detection}
	\author{A. Chiuri}
	\affiliation{Dipartimento di Fisica, Sapienza Universit\'a di Roma, I-00185 Roma, Italy}
	\author{G. Vallone}
	\affiliation{Museo Storico della Fisica e Centro Studi e Ricerche Enrico Fermi, Via Panisperna 89/A, Compendio del Viminale, I-00184 Roma, Italy}
	\affiliation{Dipartimento di Fisica, Sapienza Universit\'a di Roma, I-00185 Roma, Italy}
	\author{N. Bruno}
	\affiliation{Dipartimento di Fisica, Sapienza Universit\'a di Roma, I-00185 Roma, Italy}
	\author{C. Macchiavello}
        \affiliation{Dipartimento di Fisica ``A. Volta'' and INFN-Sezione di Pavia, via Bassi 6, 27100 Pavia, Italy}
	\author{D. Bru\ss}
        \affiliation{Institut f{\"u}r Theoretische Physik III, 
Heinrich-Heine-Universit{\"a}t D{\"u}sseldorf, D-40225 D{\"u}sseldorf, 
Germany}
	\author{P. Mataloni}
	\affiliation{Dipartimento di Fisica, Sapienza Universit\'a di Roma, I-00185 Roma, Italy}
	\affiliation{Istituto Nazionale di Ottica Applicata (INO-CNR), L.go E. Fermi 6, I-50125 Firenze, Italy}
	\date{\today}
	
	
	\begin{abstract}
We present an experimental method to produce 4-qubit phased Dicke states, 
based on a source of 2-photon hyperentangled states. By introducing quantum 
noise 
in the multipartite system in a controlled way, we have tested the robustness 
of these states. To this purpose the entanglement of the resulting multipartite 
entangled mixed states has been verified  by using a new  
kind of structural witness.
	\end{abstract}

	\maketitle

The generation and detection of multipartite entangled states is a remarkable 
challenge that needs to be accomplished in order to fully explore and exploit
the genuine quantum features of quantum information and many-body physics.
So far only a limited number of families of pure multipartite entangled states 
has been experimentally produced. In view of future applications, it is 
particularly important to test the robustness of the generated states in the 
presence of unavoidable noise coming from the environment. Here, we 
produce a new family of multipartite maximally entangled states, 
we experimentally introduce certain types of noise in a controlled way and
test the robustness properties of the states.

The experimental generation of multipartite entangled states that we
propose is based on 
hyperentangled photons \cite{barb05pra}, which allows to produce symmetric and 
phased Dicke states. 
Dicke states have recently attracted much interest, and have been
produced in experiments with photons \cite{kies07prl,prev09prl,wiec09prl}.
Phased Dicke states represent a more general family of 
entangled states with respect to the ordinary symmetric Dicke states: 
they are achieved by introducing phase changes starting from 
ordinary Dicke states.
 
In order to test the presence of multipartite entanglement we may adopt 
different kinds of entanglement witnesses. Their experimental implementations 
are presented, e.g., in Refs.~\cite{barb03prl} for bipartite 
qubits, and \cite{bour04prl,kies07prl,prev09prl,wiec09prl} for pure symmetric 
multipartite states.
In this work we implement a recently proposed new class of entanglement 
witnesses, so-called structural witnesses \cite{kram09prl}, and further 
extend such a class in order to achieve higher efficiency in entanglement detection. 
Moreover, we test the robustness of the phased Dicke states by introducing dephasing 
noise in a controlled fashion and provide a measurement of the lower 
bound on the robustness of entanglement. 
In this way we provide a new experimental
tool to investigate the entanglement properties of multipartite mixed states. 

An entanglement witness is defined as a Hermitian operator $W$ that detects 
the entanglement of a state $\rho$
if it has a negative expectation value for this state, 
$\langle W \rangle_\rho = \tr (\rho W) < 0$ while at the same time
$\tr (\sigma W) \geq 0$ for all separable states $\sigma$ 
\cite{horo96pla, terh00pla}.  
For a composite system of $N$ particles, the structural witnesses 
\cite{kram09prl} have the form
\begin{equation} \label{genentwit}
	W(k) := \id_N - \Sigma(k) \,,
\end{equation}
where $k$ is a real parameter (the wave-vector transfer in a scattering
scenario), $\id_N$ is 
the identity operator and
\begin{equation} \label{sigmagen1}
	\Sigma(k) = \frac{1}{2} [\bar\Sigma(k)+\bar\Sigma(-k)]
\end{equation}
with
\begin{eqnarray} \label{sigmagen}
	\bar\Sigma(k) &= \frac{1}{B(N,2)} 
\left( c_x \hat{S}^{xx}(k) 
+ c_y  \hat{S}^{yy}(k) + c_z \hat{S}^{zz}(k) \right), \nonumber \\
	&\quad c_i \in \R, \ |c_i| \leq 1.
\end{eqnarray}
Here $B(N,2)$ is the binomial coefficient and the structure factor operators
$\hat{S}^{\alpha\beta}(k)$ are defined as 

\begin{equation} \label{Sab}
\hat{S}^{\alpha \beta}(k) := \sum_{i<j} e^{ik(r_i -r_j)}S_i^{\alpha} S_j^\beta,
\end{equation}
where $i,j$ denote the $i$-th and $j$-th spins, $r_i, r_j$ their positions
in a one-dimensional scenario, 
and $S_i^{\alpha}$ are the spin operators with $\alpha,\beta = x,y,z$. 
In the following we normalize the distances 
with the labels of the qubits as $r_i -r_j=i-j$.
In the present work we focus on the case of 4-qubits phased Dicke 
states defined as \cite{kram09prl}:
\begin{figure*}
	\centering
		\includegraphics[width=18cm]{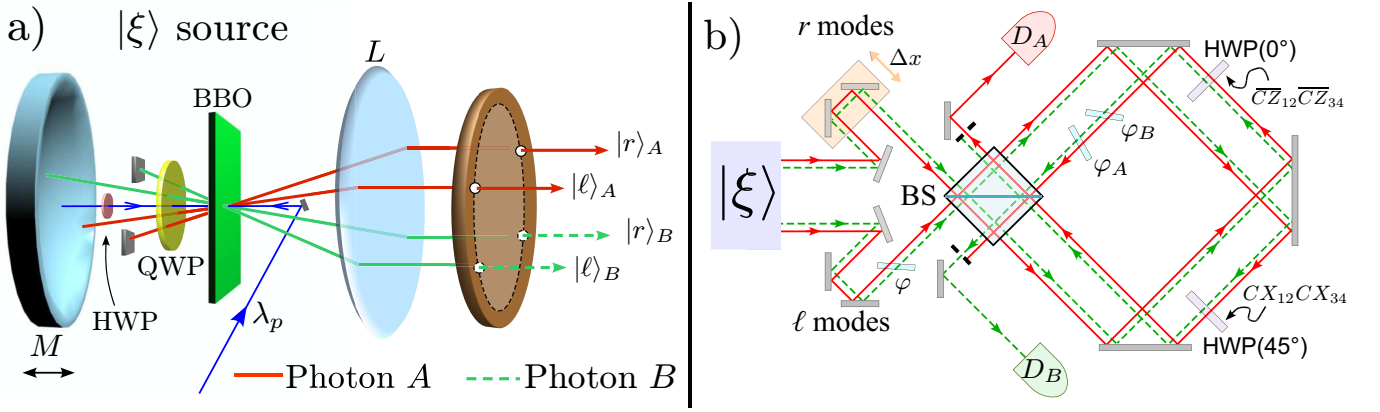}
	\caption{Generation of the 4-qubit phased Dicke state. a) Scheme of the entangled 2-photon 
	2-qubit parametric source that generates the state $\ket\xi$. 
	Mirror $M$ reflects both the UV pump beam and the parametric radiation. The lens $L$ is used to 
	obtain parallel modes at the output of the 4-hole screen. HWP (QWP) is an half- (quarter-) waveplate.
	b) Optical setup used to transform $\ket\xi$ into the state $\dicke$ and to measure Pauli operators.
	The phase $\varphi$ is used to properly generate $\ket{\xi}$ while $\varphi_{A}$ and $\varphi_{B}$ 
	are used to measure Pauli momentum operators for the A and B photon respectively.
}
	\label{fig:source}
\end{figure*}
 
\begin{eqnarray} \label{phaseddicke4}
	\ket{D_4^{ph}} = &&\frac{1}{\sqrt{6}} \big(|0011\rangle + 
|1100\rangle + |0110\rangle + |1001\rangle \nonumber \\
	&&-|0101\rangle - |1010\rangle \big).
\end{eqnarray}
A suitable structural witness for the above phased Dicke state 
is given by the operator (\ref{genentwit}) with $k=\pi$, $c_x=c_y=c_z=1$
and $S_i^{\alpha}$ being the Pauli operators \cite{kram09prl}.
This witness  leads to $\tr (\proj{D_4^{ph}} W)=-\frac49$.

A wider class of structural witness can be obtained by generalizing the operator
given in (\ref{sigmagen}) to linear superpositions of structure factor operators 
$\hat{S}^{\alpha\beta}(k)$ evaluated for different values of $k$:
\begin{eqnarray} \label{sigmagennew}
	\bar\Sigma(k^x,k^y,k^z) 
	&= \frac{1}{B(N,2)} 
\left( c_x \hat{S}^{xx}(k^x) 
+ c_y  \hat{S}^{yy}(k^y) + c_z \hat{S}^{zz}(k^z) \right), \nonumber \\
	&\quad c_i \in \R, \ |c_i| \leq 1.
\end{eqnarray}
Following the same argument as in \cite{kram09prl},
it can be 
shown  that any operator of the form (\ref{sigmagennew}) combined as in 
(\ref{genentwit}) has non-negative 
expectation values for separable states and is therefore an entanglement 
witness. 
Using this more general construction for the present experiment 
we consider a witness operator with $k^x=k^y=\pi$ and $k^z=0$:
\begin{equation} \label{wnew}
\overline W = \id_N -\frac{1}{6}\left(\hat{S}^{xx}(\pi)+\hat{S}^{yy}(\pi)-
\hat{S}^{zz}(0)\right)\ .
\end{equation}
The expectation value of the above witness for the phase Dicke state 
(\ref{phaseddicke4}) is given by $\tr (\proj{D_4^{ph}} \overline W)=-\frac23$.

{\it State generation -} We will now describe the method to generate phased
Dicke states and to implement controlled noise, and present the 
experimental results of entanglement detection. 
Let's consider the following state 
$\ket{\xi}\equiv\frac{1}{\sqrt{6}}(\ket{0010}-\ket{1000}+2\ket{0111})$.
It is easy to show that the phased Dicke state can be obtained by
applying a unitary transformation\footnote{{An equivalent and more simple 
transformation is given by $\mathcal U=Z_1CX_{12}CX_{34}H_1H_3$.
We used the transformation given in \eqref{dickexi} {
in order to compensate the optical delay introduced by the CX gates
in the Sagnac loop of Fig. \ref{fig:source}b).}}}
 $\mathcal U$ to the state $\ket \xi$:
\begin{equation}
{\ket{D^{(ph)}_{4}}=Z_4\overline{CZ}_{12}\overline{CZ}_{34}CX_{12}CX_{34}H_1H_3\ket{\xi}\equiv\mathcal U\ket{\xi}}
\label{dickexi}
\end{equation}
where $H_j$ and $Z_j$ stands for the Hadamard and the Pauli $\sigma_z$ 
transformations on qubit $j$, $CX_{ij}=\ket0_i\bra0\openone_j+\ket1_i\bra1X_j$ 
is the controlled-NOT gate and $\overline{CZ}_{ij}=\ket1_i\bra1\openone_j+\ket0_i\bra0Z_j$ 
the controlled-Z.
We realized the Dicke state by using 4-qubits encoded into
polarization and path of two parametric photons 
[A and B in figure \ref{fig:source}a)].
The $\ket{0}$ and $\ket{1}$ states are encoded into
horizontal $\ket{H}$ and vertical $\ket{V}$ polarization or into
right $\ket{r}$ and left $\ket{\ell}$ path. 
Explicitly, we used the following correspondence
between physical states and logical qubits:
\begin{align}
\label{relation}
&\{\ket{0}_{1},\ket{1}_{1}\}\rightarrow\{\ket{r}_{A},\ket{\ell}_{A}\}\\
&\{\ket{0}_{2},\ket{1}_{2}\}\rightarrow\{\ket{H}_{A},\ket{V}_{A}\}\\
&\{\ket{0}_{3},\ket{1}_{3}\}\rightarrow\{\ket{r}_{B},\ket{\ell}_{B}\}\\
&\{\ket{0}_{4},\ket{1}_{4}\}\rightarrow\{\ket{H}_{B},\ket{V}_{B}\}
\end{align}
According to these relations the state $\ket{\xi}$  reads:
\begin{equation}\label{xi}
\begin{aligned}
\ket{\xi}=\frac{1}{\sqrt{6}}[\ket{HH}(\ket{r\ell}-\ket{\ell r})+
2\ket{VV}\ket{r\ell}]
\end{aligned}
\end{equation}
and may be obtained by suitably modifying the source used to realize polarization-momentum
hyperentangled states \cite{barb05pra, cecc09prl}.
In each ``ket'' of \eqref{xi} the first (second) term refers to particle A (B). 
A vertically polarized UV laser beam impinges 
on a Type I $\beta$-barium borate (BBO) nonlinear crystal in two opposite directions, back and forth, 
and determines the generation of the polarization entangled state corresponding to the 
superposition of the spontaneous parametric down conversion (SPDC) emission 
at degenerate wavelength [see Fig. \ref{fig:source}a)]. 
A 4-hole mask selects four optical modes (two for each photon), namely $\ket{r}_A$, $\ket{\ell}_A$, 
$\ket{r}_B$ and $\ket{\ell}_B$,
within the emission cone of the crystal. 
The SPDC contribution, due to the pump beam incoming after reflection on mirror $M$, corresponds to the term 
$\ket{HH}(\ket{r\ell}-\ket{\ell r})$, whose weight is determined by a
half waveplate intercepting the UV beam (see \cite{vall07pra} for more details on the generation of 
the non-maximally polarization entangled state).
The other SPDC contribution $2\ket{VV}\ket{r\ell}$ is determined
by the first excitation of the pump beam: 
here the $\ket{\ell r}$ modes are intercepted by two beam stops and a quarter waveplate QWP transforms the $\ket{HH}$ SPDC emission into $\ket{VV}$ after reflection on mirror $M$. 
The relative phase between the $\ket{VV}$ and $\ket{HH}$ is varied by
translation of the spherical mirror.

The transformation \eqref{dickexi} $\ket{\xi}\rightarrow\ket{D^{(ph)}_{4}}$ 
is realized by usign waveplates and one beam splitter (BS): 
the two Hadamards $H_1$ and $H_3$ in \eqref{dickexi},
acting on both path qubits, are implemented by a single BS
for both A and B modes. For each controlled-NOT (or controlled-Z) gate appearing 
in \eqref{dickexi} the control and target qubit
are respectively represented by the path and the polarization of a single photon:
a half waveplave (HWP) with axis oriented at 45$^\circ$ (0$^\circ$) with respect to the vertical direction
and located into the left $\ket{\ell}$ (right $\ket r$) mode implements a CX ($\overline{CZ}$) gate.

{After these transformations, 
the optical modes are spatially matched for a second time on the BS, 
closing in this way a ``displaced Sagnac loop'' interferometer that allows
high stability in the path Pauli operator measurements [see Fig. \ref{fig:source}b)].
Polarization Pauli operators are measured by standard polarization analysis setup 
in front of detectors $D_A$ and $D_B$ (not shown in the figure).
Note that, the $\ket{0}$ ($\ket{1}$) states
are identified by the counterclockwise (clockwise) modes in the Sagnac loop.}

\begin{figure}[t]
	\centering
		\includegraphics[width=8.5cm]{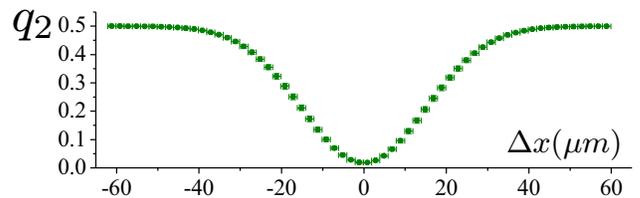}
	\caption{Values of $q_2$ corresponding to different values of the path delay $\Delta x$.}
	\label{fig:q2}
\end{figure}

{\it Decoherence -} We will 
now describe how we introduced a controlled decoherence into the system
(we mention that recently controlled decoherence has been implemented in an
ion trap experiment \cite{barr10qph}).
Consider a single photon in a Mach-Zehnder interferometer with two arms (left and right).
Varying the relative delay $\Delta x=\ell-r$ between the right and left arm 
corresponds to a single qubit path decoherence channel given by 
$\rho\rightarrow(1-p)\rho+{p}Z\rho Z$. 
The parameter $p$ is related to $\Delta x$:
when $\Delta x>\tau$, where $\tau$ represents the photon coherence time, 
then $p=\frac12$, while when $\Delta x=0$ we have $p=0$. 
This can be understood by observing that there are two time bins (one for each path).
By varying the optical delay, we entangle the path with
the time bin degree of freedom (DOF). Hence, by tracing over time we obtain decoherence in the
path DOF depending on the overlap between the two time bins.
In our setup, this can be obtained by changing 
the relative delay $\Delta x=\ell-r$ between the right and the left modes of the photons in the
first interferometer shown in Fig. \ref{fig:source}.
Since the translation stage acts simultaneously on both photons, this operation
corresponds to two path decoherence channels:
\beq
\label{decxi}
\rho\rightarrow(1-q_2)^2\rho+q_2(1-q_2)\left[Z_1\rho Z_1+Z_3\rho Z_3\right]+q^2_2Z_1Z_3\rho Z_1Z_3
\eeq
where the parameter $q_2$ is related to $\Delta x$ in the following way. 
Let's consider the path terms in the 
$\ket{HH}$ contribution in $\ket\xi$, namely $\ket{\psi^-}=\frac{1}{\sqrt{2}}(\ket{r\ell}-\ket{\ell r})$.
The decoherence acts by (partially) spoiling the coherence between the $\ket{r\ell}$ and $\ket{\ell r}$ term
giving the state $\frac12(\ket{\ell r}\bra{\ell r}+\ket{r\ell}\bra{r\ell})-\frac12(1-2q_2)^2
(\ket{\ell r}\bra{r\ell}+\ket{r\ell}\bra{\ell r})$. 
By assuming that for $\ket{\psi^-}$ the decoherence \eqref{decxi} is the main source of imperfections, 
the measured visibility\footnote{The measured 
visibility is defined as $\widetilde V_{exp}(\Delta x)=\frac{B-C}{B}$
where $B$ are the coincidences measured out of interference 
{(i.e. measured for $\Delta x$ much longer than the single photon coherence
lenght)} and $C$ the coincidences measured
in a given position of $\Delta x$.} 
$\widetilde V_{exp}(\Delta x)$
of first interference on BS may be compared with the calculated value $\widetilde V=(1-2q_2)^2$:
then, the relation between $\Delta x$ and $q_2$, shown in Fig. \ref{fig:q2}, is obtained. 
It is worth noting that at $\Delta x=0$ we have
$q_2=0.0175\pm0.0001$ which corresponds to a maximum visibility $V_{exp}=0.9313\pm0.0005$
at $\Delta x=0$.

The decoherence channel \eqref{decxi} acts on the state $\ket{\xi}$. 
However, it can be interpreted as a decoherence
acting on the phased Dicke state $\dicke$.
Using equation \eqref{dickexi} and the relations {
$\mathcal UZ_1\mathcal U^\dag=-Y_1Y_2$ and 
$\mathcal UZ_3\mathcal U^\dag=Y_3Y_4$}, the channel \eqref{decxi} may be interpreted as a
{\it collective decoherence channel} on $\dicke$:
\beq
\label{dec2}
\dicke\bradicke\rightarrow \sum^4_{j=1}B_j\dicke\bradicke B^\dag_j
\eeq
with
{$B_1=(1-q_2)\openone$, 
$B_2=\sqrt{q_2(1-q_2)}Y_1Y_2$, $B_3=\sqrt{q_2(1-q_2)}Y_3Y_4$
and $B_4=q_2Y_1Y_2Y_3Y_4$.} 
A collective decoherence is a decoherence process that cannot be seen
as the action of several channels acting separately on two (or more) qubits.
A different type of collective noise, introduced in \cite{macc02pra}, was
experimentally demonstrated in \cite{bana04prl} for two polarization qubits in
optical fibers.

Two other main sources of imperfections must be considered in our setup 
{(see supplementary informations for
a detailed discussion)}: the first one is due to a non perfect superposition between forward 
and backward SPDC emission, i.e. between the $\ket{HH}$ and $\ket{VV}$ contributions.
This imperfection can be modeled as a phase polarization decoherence channel acting on qubit $2$:
$\rho\rightarrow(1-q_1)\rho+q_1 Z_2\rho Z_2$. By selecting in $\ket\xi$ the correlated modes 
$\ket{rl}$ and by suitably setting the HWP on the pump beam we obtain the following state:
$\frac{1}{\sqrt{2}}(\ket{HH}_{AB}+e^{i\gamma}\ket{VV}_{AB})\ket{rl}$. 
Even in this case the value of the measured polarization visibility ($V_\pi\simeq 0.90$)
can be related to the polarization decoherence channel as 
$q_1=\frac{1-V_\pi}{2}\simeq0.05$. 
{The second interference on the BS (i.e. after the Sagnac loop)} has been also investigated. 
In the measurement condition we obtained 
an average visibility of $V_{k_2}\simeq0.80$ corresponding to a decoherence channel
$\rho\rightarrow(1-q_3)^2\rho+q_3(1-q_3)\left[Z_1\rho Z_1+Z_3\rho Z_3\right]+q^2_3Z_1Z_3\rho Z_1Z_3$
with $q_3=0.05$.

\begin{figure}
	\centering
		\includegraphics[width=9cm]{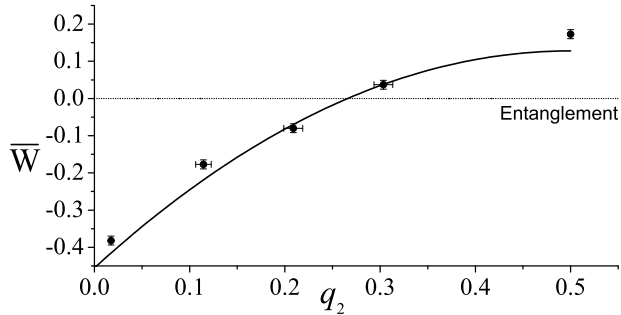}
	\caption{Experimental values of the witness $\overline W$ as a function of $q_2$. 
	The dark curve corresponds to the theoretical curve obtained by 
setting $q_1=0.05$ and $q_3=0.05$.}
	\label{fig:W}
\end{figure}

{\it Measurements -} 
{We measured the witness operator \eqref{wnew} for different values of $q_2$. 
The results are shown in figure \ref{fig:W}. 
The dark curve corresponds to the theoretical curve obtained by 
considering all the three described 
decoherence channels and setting $q_1=0.05$ and $q_3=0.05$ (see 
the supplementary informations about the details on the theoretical curve).
Notice that the noise parameter for which the witness expectation value
vanishes gives a lower bound on the robustness of the entanglement of the
produced state with respect to the implemented noise.
The witness $\overline W$ measured for the phased Dicke state is 
\beq\label{Wexp}
\langle{\overline W}\rangle_{exp}=-0.382\pm0.012
\eeq
}

{We also measured a witness $W_{mult}$ introduced in \cite{toth09njp}
to demonstrate that the generated 
state $\dicke$ is a genuine multipartite state and to obtain a bound on the fidelity $F$. 
Its expression is given in the supplementary informations.
We obtained 
\beq
\langle W_{mult}\rangle=-0.341\pm0.015\quad\rightarrow\quad F\geq0.780\pm0.005
\eeq
}

Following the approach of quantitative entanglement witnesses \cite{eise07njp},
we can also use the experimental result on the expectation value of the witness
to provide a lower bound on the {\it random robustness} of entanglement $E_r$.
This is defined in \cite{vida99pra} to be the maximum amount of white noise that
one can add to a given state $\rho$ before it becomes separable. 
A lower bound on $E_r(\rho)$ is given by

\begin{equation}
E_r(\rho) \ge \frac{D|\tr (\rho \overline W)|}{\tr (\overline W)}\;,
\label{Er}
\end{equation}
where $D$ is the dimension of the Hilbert space on which $\rho$ acts.
In our experiment the witness from Eq. (\ref{wnew}) and its expectation value 
given in Eq. (\ref{Wexp}) lead to

{\begin{equation}
\label{Erexp}
E_r(\rho) \ge |\langle{\overline W}\rangle_{exp}|=0.382\pm0.012
\end{equation}
}

In summary, we have generated phased 4-qubit Dicke states
with hyperentangled photons. We demonstrated the implementation of controlled 
noise via a relative path delay in the interferometer.
The multipartite entanglement was detected via a new class of
structural entanglement witnesses and the robustness of entanglement was
tested by using an intrinsically high phase stability setup.
The realized phase Dicke states have a high fidelity and, compared with other Dicke states based on 4-photon entanglement, are produced at higher repetition rate. 

This work was supported in part by the EU projects CORNER, DFG project
and by Finanziamento Ateneo 2009 of Sapienza Universit\`a di Roma.


\newpage\ \newpage
\section{SUPPLEMENTARY INFORMATION}
Let's now describe in more detail the considered decoherence sources. 
First of all, in our setup, there is a polarization decoherence 
at the level of the $\ket\xi$ generation due to a non perfect superposition between the
$\ket{HH}$ and $\ket{VV}$ emission.
Since $\ket\xi$ is given by the superposition of $\ket{HH}$ and $\ket{VV}$ terms, 
our decoherence partially erases 
the coherence between them but cannot introduce terms containing $\ket{VH}$ or $\ket{HV}$. 
This decoherence can be modeled by a phase decoherence channel acting on polarization qubit $2$:
\beq\label{dec_pi}
\rho\rightarrow(1-q_1)\rho+q_1 Z_2\rho Z_2
\eeq
By exploiting the same arguments used to obtain \eqref{dec2}, 
the channel \eqref{dec_pi} can be interpreted as a
decoherence channel on $\dicke$. Since $\mathcal UZ_2\mathcal U^\dag=Z_1Z_2$,
the polarization decoherence \eqref{dec_pi} can be written as
\beq
\dicke\bradicke\rightarrow \sum^2_{j=1}A_j\dicke\bradicke A^\dag_j
\eeq
with $A_1=\sqrt{1-q_1}\openone$ and $A_2 =\sqrt{q_1}Z_1Z_2$.
By measuring the visibility of polarization interference we estimated $q_1\simeq0.05$.

The second decoherence affects the path degree of freedom and corresponds to the channel given in eq.
\eqref{decxi}. We can change the parameter $q_2$ by varying the delay $\Delta x$ in the
first interferometer. If figure \ref{fig:q2} we show the relation between the parameter $q_2$ and
the path delay $\Delta x$.

A third decoherence effect, again in the path degree of freedom, is related to the second
interference on the BS. The non-perfect interference can be modeled as a decoherence
channel acting exactly as \eqref{decxi}. Written in the Kraus representation
it reads:
\beq
\dicke\bradicke\rightarrow \sum^4_{k=1}C_k\dicke\bradicke C^\dag_k
\eeq
with $C_1=(1-q_3)\openone$, $C_2=\sqrt{q_3(1-q_3)}Z_1$, $C_3=\sqrt{q_3(1-q_3)}Z_3$ and
$C_4=q_3Z_1Z_3$.
By measuring the interference visibility we estimated $q_3\simeq0.05$.

The three decoherence channels can be summarized as follows
\beq
\rho(q_1,q_2,q_3)\equiv
\sum^4_{k=1}\sum^4_{j=1}\sum^2_{i=1}C_kB_jA_i\dicke\bradicke A^\dag_i B^\dag_jC^\dag_k
\eeq
From the previous expression, it is possible to calculate the theoretical expectation 
values of the operators appearing in the witness as a function of the $q$'s parameters: 
\eqref{wnew}:
\beq
\begin{aligned}
\langle S_{xx}(\pi)\rangle=&4-\frac83q_3(3-q_3)\\
&-\frac{16}3(1-q_3)^2[q_1(1-2q_2)^2+2q_2(1-q_2)]\\
\langle S_{yy}(\pi)\rangle=&4-\frac{16}{3}q_{1}(1-q_{3})^{2}+\frac83(q_{3}-3)q_{3}]\\
\langle S_{zz}(0)\rangle=&-2+\frac{16}3q_2(1-q_2)\\
\end{aligned}
\eeq
{For $q_1=0.05$ and $q_3=0.05$ we obtain the following expectation value for
$\overline W$:
\beq
\langle \overline W\rangle=-0.455+2.333\;q_2-2.333\;q^2_2
\eeq
This expression is used for the theoretical curve in figure \ref{fig:W}.

The witness used to detect multipartite entanglement (see eq. (36) of \cite{toth09njp})
is
\begin{widetext}
\beq
W_{mult}=\frac18\left[21-2S_{xx}(\pi)-2S_{yy}(\pi)+S_{zz}(0)-2X_1X_2X_3X_4-2Y_1Y_2Y_3Y_4-
7Z_1Z_2Z_3Z_4\right]
\eeq
\end{widetext}
By following \cite{toth09njp} it is possible to obtain a bound for the fidelity:
\beq
F > \frac23-\frac13\langle W_{mult}\rangle.
\eeq
}

\end{document}